\documentclass{article}

\usepackage[preprint]{neurips_2024}

\usepackage[utf8]{inputenc} 
\usepackage[T1]{fontenc}    
\usepackage{hyperref}       
\usepackage{url}            
\usepackage{booktabs}       
\usepackage{amsfonts}       
\usepackage{nicefrac}       
\usepackage{microtype}      
\usepackage{xcolor}         

\usepackage{soul}
\usepackage{graphicx}
\usepackage{booktabs}
\usepackage{multirow}
\usepackage[normalem]{ulem}
\useunder{\uline}{\ul}{}
\usepackage{amsmath}
\usepackage{float}

\title{DSPO: An End-to-End Framework for Direct Sorted Portfolio Construction}

\author{%
    Jianyuan Zhong\textsuperscript{1}\thanks{These authors contributed equally to this work.}, 
    Zhijian Xu\textsuperscript{1}\footnotemark[1], 
    Saizhuo Wang\textsuperscript{2}\thanks{Work during an internship at IDEA Research.}, 
    Xiangyu Wen\textsuperscript{1}, 
    Jian Guo\textsuperscript{3}\thanks{Corresponding author}, 
    Qiang Xu\textsuperscript{1}\footnotemark[3] \\
    \textsuperscript{1}The Chinese University of Hong Kong \\
    \textsuperscript{2}The Hong Kong University of Science and Technology \\
    \textsuperscript{3}IDEA Research \\
    \texttt{\{jyzhong, zjxu21, xywen22, qxu\}@cse.cuhk.edu.hk}, \\
    \texttt{swangeh@connect.ust.hk}, \texttt{guojian@idea.edu.cn}
}

\begin{document}

\maketitle

\vspace{-15pt}
\begin{abstract}

\vspace{-5pt}
In quantitative investment, constructing characteristic-sorted portfolios is a crucial strategy for asset allocation.
Traditional methods transform raw stock data of varying frequencies into predictive characteristic factors for asset sorting, often requiring extensive manual design and misalignment between prediction and optimization goals.
To address these challenges, we introduce Direct Sorted Portfolio Optimization (DSPO), an innovative end-to-end framework that efficiently processes raw stock data to construct sorted portfolios directly.
DSPO's neural network architecture seamlessly transitions stock data from input to output while effectively modeling the intra-dependency of time-steps and inter-dependency among all tradable stocks. Additionally, we incorporate a novel Monotonical Logistic Regression loss, which directly maximizes the likelihood of constructing optimal sorted portfolios. To the best of our knowledge, DSPO is the first method capable of handling market cross-sections with thousands of tradable stocks fully end-to-end from raw multi-frequency data. Empirical results demonstrate DSPO's effectiveness, yielding a RankIC\footnote{RankIC, a non-parametric rank-based measure of prediction accuracy, measures the Spearman correlation between predicted and actual asset returns}  of 10.12\% and an accumulated return of 121.94\% on the New York Stock Exchange in 2023-2024, and a RankIC of 9.11\% with a return of 108.74\% in other markets during 2021-2022.
\end{abstract}

\vspace{-10pt}

\begin{figure}[h]
    \centering
    \includegraphics[width=0.95\textwidth]{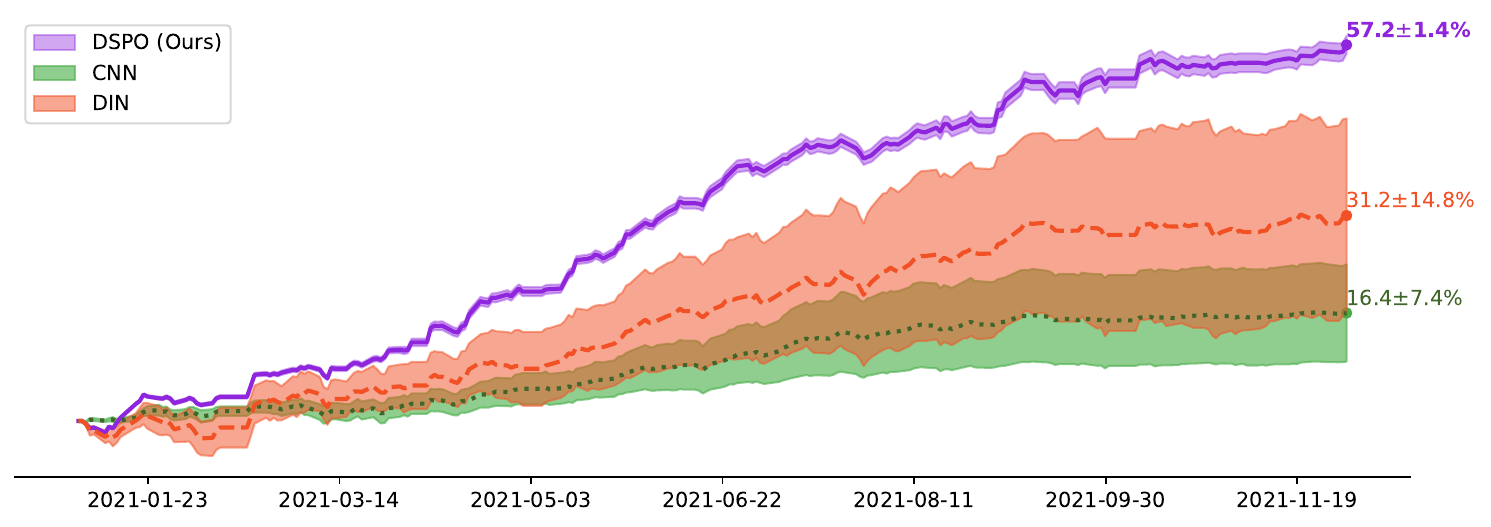}
    \vspace{-10pt}
    \caption{Comparison of the mean accumulative return (colored lines) and its variance across eight trials (color spread around the lines) during the 2021-2022 backtesting period in A-Share market. Our \textbf{Direct Sorted Portfolio Optimization (DSPO)} framework \textbf{DOUBLES} the accumulative return with \textbf{1/10} the variance across different trials over other approaches.}
    \label{fig:mesh1}
\end{figure}

\section{Introduction}
\label{intro}
\vspace{-2pt}
Portfolio construction is a pivotal challenge in quantitative investment, where the goal is to allocate capital across a diverse asset pool to maximize profits. A well-regarded method for achieving this objective is the characteristic-sorted portfolio, which involves sorting assets based on specific traits before capital allocation~\cite{cattaneo2020characteristic}. This approach has attracted significant attention because of its potential for high returns and its critical role in successful investment strategies.

Current methods typically adhere to a pipelined process that begins with feature engineering \cite{kakushadze_101_2016, zhang2020autoalpha, lin2022factors, alphagpt2023}, followed by modeling~\cite{jiang_applications_2021, hu_survey_2021}, and concludes with portfolio sorting~\cite{cattaneo2020characteristic}. Despite its widespread adoption due to practicality, this approach encompasses several limitations. Initially, feature engineering demands significant expertise and effort to extract predictive features that are critical for the success of subsequent stages in the pipeline. Besides, these features often becomes outdated due to the dynamic nature of financial markets~\cite{shin2024synergistic}. More critically, there is an inherent mismatch between the objectives of asset prediction and portfolio optimization~\cite{elmachtoub_smart_2022}, which frequently results in suboptimal performance.

End-to-end learning offers a promising solution, where a neural network seamlessly processes inputs to outputs, adapting to changes in data distribution through retraining or fine-tuning~\cite{hoi2021online} and optimizing directly for specific goals. However, the application of end-to-end learning to sorted portfolio construction presents significant challenges, primarily in optimization. Models are typically assessed based on their ability to accurately rank assets by expected returns, a task complicated by the non-differentiability of sorting functions. Alternative optimization strategies such as reinforcement learning~\cite{ijcai2023p530, wang_deeptrader_2021}, have been explored but often demonstrate training instability \cite{sun2023trademaster, sun_prudex-compass_2022} that affects consistency and reliability. While some efforts have processed uni-frequency raw stock data end-to-end, they have not demonstrated effectiveness across market cross-sections containing multi-frequency data for thousands of tradable stocks~\cite{liu2023deep}.


To address these challenges, we introduce the Direct Sorted Portfolio Optimization (DSPO), an end-to-end, deep learning-based framework specifically designed for constructing sorted portfolios across entire market cross-sections. DSPO incorporates several innovative components tailored to address distinct challenges. The Stock-wise Multi-Frequency Fusion Module leverages Convolutional Neural Networks (CNNs)~\cite{he2016deep} and Transformers~\cite{vaswani2017attention} to process data at varying frequencies, thereby integrating multi-frequency inputs into a unified representation. Additionally, the Inter-Stock Transformer captures the interdependencies among all tradable assets in the market using attention mechanisms. To address the non-differentiable nature of the ranking objective, we introduce the Monotonical Logistic Regression (MonLR) loss, which directly predicts the likelihood of constructing optimal sorted portfolios. To further reduce overfitting and training instability, which are exacerbated by limited sample sizes from cross-sectional data, we implement a sub-sampling strategy. This approach not only increases the number of unique samples but also balances the trade-off between estimation accuracy and sample size.

To validate the effectiveness of the proposed DSPO model, we conduct comprehensive experiments using recent data from two of the largest markets, New York Stock Exchange (NYSE) and China A-shares market (A-Share). Our experimental framework centers on intra-day trading, forming sorted portfolios from over 4,000 stocks by selecting the top and bottom 10\% for long and short positions, respectively. The results underscored DSPO’s exceptional performance, achieving a RankIC of 10.12\% and an accumulated return of 121.94\% on the NYSE, alongside a RankIC of 9.11\% and an accumulated return of 108.74\% in the A-Share market. Further, our ablation studies on DSPO’s modules confirm the significance of each component, while experiments on sub-sampling strategies demonstrate that empirical losses in our training procedure still converge to the expected outcomes when directly optimizing for sorting. We also conduct multiple case studies to showcase DSPO's adaptability to evolving markets with deteriorating factors, achieving significantly lower variance across different runs. These results not only highlight DSPO's superior capacity to leverage diverse market data but also its robustness in managing the complexities of modern financial markets, markedly surpassing traditional benchmark methods.
\section{Related Works and Motivation}
\label{related_work}



\paragraph{Feature Engineering in Traditional Methods}
Conventional pipelined approaches to portfolio construction heavily depend on feature engineering to harmonize raw financial data from multiple sources with varying update frequencies (e.g., monthly/weekly fundamental data versus daily volume-price data versus intraday quotes and transactions) for actionable insights\cite{zhang2020autoalpha, lin2022factors}. Historically, extracting relevant features has required deep domain knowledge, including a comprehensive understanding of both technical and fundamental indicators that support predictive models in forecasting stock performance\cite{alphagpt2023}. However, these engineered features tend to degrade as markets evolve, leading to increasingly labor-intensive efforts to discover new features. While techniques such as PCA and LASSO\cite{chen2022multi, yun2022prediction} have been employed to address these challenges through dimensionality reduction and regularization, the fundamental limitations of relying on manually curated features have spurred interest in more autonomous learning frameworks.

\paragraph{Deep Learning Applications in Quantitative Finance}
Deep learning has revolutionized quantitative finance by enhancing traditional financial modeling with advanced predictive capabilities. Convolutional Neural Networks (CNNs) have improved forecasting accuracy by analyzing patterns in financial time series data \cite{8920761, chen2022multi, Borovkova2019AnEO, Fischer2017DeepLW}. Additionally, Recurrent Neural Networks (RNNs), particularly Long Short-Term Memory (LSTM) networks, effectively capture temporal dependencies in stock price movements \cite{chen2022multi, Borovkova2019AnEO, Fischer2017DeepLW}. Other advancements include the introduction of Transformer\cite{vaswani2017attention} architectures and graph neural networks \cite{sawhney_spatiotemporal_2020, sawhney_stock_2021} that model inter-stock or intra-timestep dependencies \cite{wang2022mtmd, yoo_soun_park_kang_2021}.
However, these approaches, while innovative, often employ regression~\cite{wang2022mtmd, yoo_soun_park_kang_2021} and classification as objectives~\cite{9154061}, which focus on predicting specific asset features, generally do not optimize directly for ranking accuracy.
Alternative optimization strategies such as reinforcement learning~\cite{ijcai2023p530, wang_deeptrader_2021} have been explored but often demonstrate training instability \cite{sun2023trademaster, sun_prudex-compass_2022} that affects consistency and reliability. While certain studies~\cite{liu2023deep} have employed end-to-end processing of raw stock data with a focus on optimizing the Sharpe ratio, they may not fully demonstrate transferability or scalability when applied to a broader market cross-section, which often includes thousands of tradable stocks.


\paragraph{Deep Learning for Ranking}
Deep learning has proven highly effective in ranking applications within recommendation systems and information retrieval. Techniques like Restricted Boltzmann Machines (RBMs), RankNet, and DeepRank have significantly enhanced our understanding of user preferences and query-document relationships, showcasing the prowess of deep learning in handling complex data structures \cite{rendle2012deep, xia2013listwise, guo2015deep}. Notably, some studies \cite{feng2019temporal, sawhney_stock_2021} employ pairwise ranking loss alongside for stock prediction. However, these method have several limitations. Pairwise ranking loss is not a maximum likelihood estimator, necessitating the inclusion of all data from an entire market cross-section to accurately estimate the ranking relationship between predicted and actual returns. This requirement results in high memory demands, especially when handling multi-frequency data.
Moreover, learning from the entire cross-section substantially reduces the number of available training samples compared to other methods, increasing the risk of overfitting.
Additionally, this type of loss can introduce numerical instability, further complicating the training process.

To this end, we redefine the problem of constructing sorted portfolios by directly optimizing the likelihood that the predicted future returns preserve the ranking order of the actual returns. Our objective function acts as a maximum likelihood estimator, benefiting from learning diverse probability distributions as outlined in the literature \cite{oord2016wavenet, van2017neural, van2016conditional}. Additionally, a maximum likelihood estimator enables learning from sub-sampled market cross-sections, significantly increasing the diversity and number of training samples and reducing memory demands. Accordingly, we develop an end-to-end framework, with a neural network architecture capable of directly optimizing sorted portfolios from multi-frequency input, thus removing the need for extensive feature engineering and maintaining good results across different runs.

\section{Methodology}
\label{method}



\subsection{Problem Formulation}
The input consists of high-frequency market data and daily-frequency fundamental data for actively traded stocks, represented as \(X = \{x_1, x_2, \ldots, x_n\}\). Each data point \(x_i\) includes both high-frequency (\(x_i^{hf}\)) and low-frequency (\(x_i^{lf}\)) components. The high-frequency component, \(x_i^{hf} \in \mathbb{R}^{T \times a}\), encompasses market data elements such as minute-frequency prices (high, low, open, close) and the ask-bid spread. The low-frequency component, \(x_i^{lf} \in \mathbb{R}^b\), contains fundamental data attributes like P/E ratio, P/B ratio, PEG ratio, and turnover metrics, where \(a\) and \(b\) denote the dimensions of the high-frequency and low-frequency data vectors, respectively.

Our core computational model is a neural network parametrized by \(\theta\), denoted as \(f(\cdot; \theta)\). This function processes the concatenated feature sets from both frequency domains to predict a score, \(S \in \mathbb{R}^N\), where \(N\) is the number of active stocks. The predicted scores aim to preserve the sorted order of the ground truth future returns \(S = f(X; \theta)\).

\subsection{Model Architecture}
\begin{figure}
    \centering
    \makebox[\textwidth][c]{\includegraphics[width=1.0\textwidth]{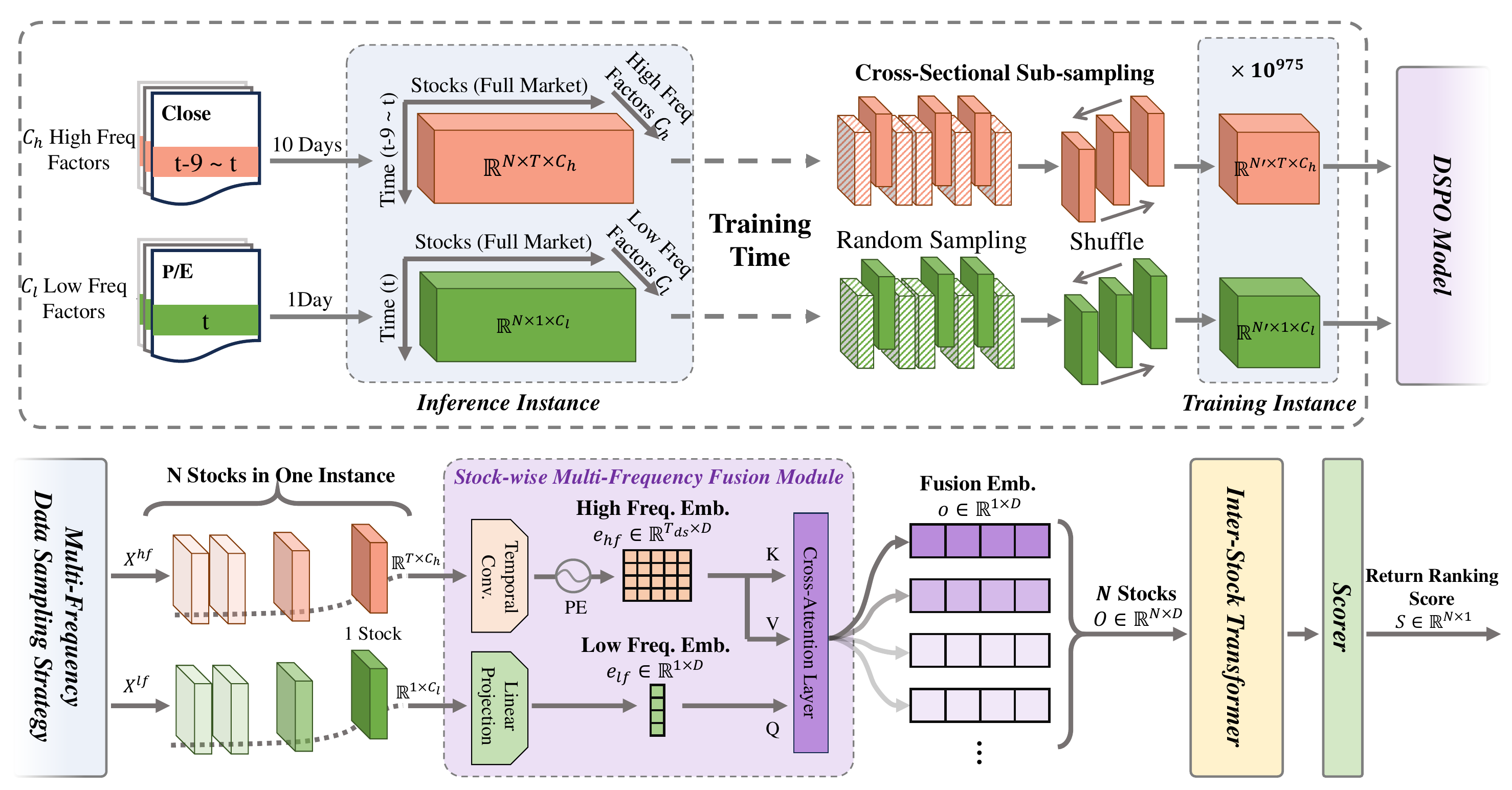}}
    \caption{\textbf{Upper Panel: data sampling strategy.} For inference, we input the high-freq. data and daily low-freq. data of the entire market cross-section. For training, we perform cross-section sub-sampling. \textbf{Lower Panel: neural network architecture.} Stock-wise multi-frequency fusion module integrates high-freq. and low-freq. data to a unified fusion embedding. Inter-Stock Transformer correlates all stock-wise fusion embeddings, subsequently outputting a ranking score with an MLP scorer. }
    \label{fig:model}
\end{figure}

As illustrated in Fig.~\ref{fig:model}, Our model architecture is specifically designed to address the problem formulation by processing both high-frequency and low-frequency data inputs to directly output a score indicating the strength of predicted future returns. The architecture comprises two main modules:

\paragraph{Stock-wise Multi-Frequency Fusion Module}
This module is engineered to capture and integrate the multi-frequency temporal dynamics of each stock into a cohesive representation. High-frequency data undergo convolutional operations to delineate temporal relationships, followed by a linear projection into an intermediate high-frequency embedding space (\( e_i^{hf} \in \mathbb{R}^{T \times D} \)), where \( D \) represents the dimension of the shared embedding space. Concurrently, low-frequency data are linearly projected into the same embedding dimension (\( e_i^{lf} \in \mathbb{R}^{1 \times D} \)), facilitating a unified fusion through a cross-attention mechanism within a Transformer architecture:

\begin{equation}
    o_i = \text{Softmax}\left(\frac{(e_i^{lf} W^{stock}_q) (e_i^{hf} W^{stock}_k)^\top}{\sqrt{D}}\right) (e_i^{hf} W^{stock}_v)
\end{equation}

This operation outputs \(o_i \in \mathbb{R}^{1 \times D}\), which will then be used to compute a unified representation that integrates information from both data frequencies.

\paragraph{Inter-Stock Transformer Module}
Designed to model the interdependencies among all tradable assets within the same market cross-section, this module utilizes the concatenated representations \(O = [o_1, \ldots, o_N] \in \mathbb{R}^{N \times D}\). The interdependencies are modeled using self-attention:

\begin{equation}
    R = \text{Softmax}\left(\frac{(O W^{intra}_q) (O W^{intra}_k)^\top}{\sqrt{D}}\right) (O W^{intra}_v)
\end{equation}

The resulting ranking representation \(R = [o_1, \ldots, o_N] \in \mathbb{R}^{N \times D}\) is subsequently mapped to scores representing the ranking of future returns through a multilayer perceptron (MLP):

\subsection{Monotonic Logistic Regression Loss for Stable Optimization}
\label{sec:monLR}
\paragraph{Derivation of Monotonic Logistic Regression Loss (MonLR)}
We address a ranking problem by correlating predicted returns \(X = (x_1, x_2, \dots, x_n)\) with actual returns \(Y = (y_1, y_2, \dots, y_n)\). The objective is to maximize the likelihood that a monotonic function describes the relationship between \(X\) and \(Y\), facilitating the construction of an optimally sorted portfolio. This task is equivalent to maximizing the probability that the predicted relative ordering of returns matches the actual ordering. This can be expressed as maximizing the probability that the signs of the differences in returns are correctly predicted:
\begin{equation}
    \theta^* := \underset{\theta}{\text{argmax}} \prod_{i=1}^{N} \prod_{j=1}^{N} \Pr(\operatorname{sign}(y_i - y_j) = \operatorname{sign}(f(x_i; \theta) - f(x_j; \theta)))
\end{equation}

The \(\operatorname{sign}\) function indicates the direction of the difference between two numbers. To enhance optimization and numerical stability, while minimizing the negative log-likelihood (logistic loss), we employ the hyperbolic tangent function (\(\operatorname{tanh}\)) as a smoother alternative to the sign function. The optimization objective then becomes:
\begin{equation}
\underset{\theta}{\text{argmin}} \, -\log(\mathcal{L}(\theta)) = \sum_{i=1}^N \sum_{j=1}^N \log\left(1 + \exp\left(-\operatorname{tanh}(f(x_i; \theta) - f(x_j; \theta)) \cdot \operatorname{tanh}(y_i - y_j)\right)\right)
\end{equation}

To adapt this formulation to a practical machine learning scenario, we define the expected loss over the dataset, reflecting the expected performance across different market cross-sections:

\begin{equation}
    \mathbb{E}_{(X, Y) \sim \mathcal{D}}\left[\sum_{i=1}^N \sum_{j=1}^N \log\left(1 + \exp\left(-\operatorname{tanh}(f(x_i; \theta) - f(x_j; \theta))\right) \cdot \operatorname{tanh}(y_i - y_j)\right)\right]
    \label{eq:all}
\end{equation}

\paragraph{Cross-Sectional Sub-Sampling}
Eq.~\ref{eq:all} requires the entire market cross-section to form one training sample, which typically yields fewer samples than in other settings (e.g., 756 samples for a 3-year intra-day trading dataset). This scarcity can undermine training stability and lead to overfitting. To address this, we propose an \textit{Cross-Section Sub-Sampling Strategy}. We randomly select \(m\) distinct trading days from our dataset to form \(m\) cross-sections, then randomly choose \(k\) stocks from these cross-sections to create a training mini-batch of \(m\) samples.

The reformulated problem using our sub-sampling strategy is expressed as:

\begin{equation}
    \hat{\mathbb{E}}_{(X', Y') \subset (X, Y) \sim \mathcal{D}}\left[\sum_{i=1}^k \sum_{j=1}^k \log\left(1 + \exp\left(-\operatorname{tanh}(f(x_i; \theta) - f(x_j; \theta))\right) \cdot \operatorname{tanh}(y_i - y_j)\right)\right]
    \label{eq:emp}
\end{equation}

Where \((X', Y')\) is the sample obtained from our subsampling strategy. For a 3-year intra-day trading dataset, our method can yield about \(10^{975}\) unique \((X', Y')\)s. Intuitively, setting \(k\) large enough to reflect the characteristics of our market cross-sections and training over a sufficient number of iterations should allow the expectation Eq.~\ref{eq:emp} of this empirical loss to converge to that of Eq.~\ref{eq:all}, for optimizing directly on the entire cross-section. Empirical results supporting this strategy's effectiveness in preventing overfitting and achieving comparable performance with varying \(k\) values will be presented in the following section.

\section{Experiments}
\label{experiments}

\subsection{Experimental Settings}
To demonstrate the effectiveness of DSPO, we conduct experiments on recent data from the two largest markets: NYSE and A-Share. We focus on a cross-section intra-day trading scenario, where models construct a sorted portfolio at market close on day \(T\) and execute transactions at market open on day \(T+1\). To create a sorted portfolio, we select the top 10\% and bottom 10\% of over cross-sections with over 4,000 stocks for long and short positions, respectively.

\paragraph{Model Training Strategy} For the New York Stock Exchange (NYSE), we utilize transaction data spanning 2020 to 2022 for model training, with evaluations conducted in 2023 using a rolling retraining approach every three months. In the case of the A-Share market, training is performed using data from 2018 to 2020, followed by direct deployment in 2021. Importantly, the selection of these particular datasets and time frames \textit{\textbf{is not intended to}} artificially enhance performance relative to other methods. Instead, this choice is dictated by the availability and economic feasibility of acquiring comprehensive high-frequency data, which is both costly and challenging to transfer in its raw form. These datasets represent the most recent and accessible data we were able to procure for both markets.

\paragraph{Data Collection \& Preliminary Filtering} We collect recent data from the NYSE through polygon.io\footnote{Polygon.io is a financial data platform that provides both real-time and historical market data, which allow their data for education, research and personal use.} and obtained A-Share data through professional contacts. Both datasets include high-frequency trading signals (e.g., open, high, low, close, volume) and low-frequency data extracted from financial reports (e.g., price-to-book ratio, price-to-earning-growth ratio, turnover rates). We applied standard normalization to high-frequency data per stock and MinMax normalization to low-frequency data, ranging from 0 to 1. The dataset for each stock comprises 10 days of high-frequency data and 1 day of low-frequency data. We excluded stocks with no trading volume over the 10-day period due to data noise.

\paragraph{Baseline Formulations} In order to assess the performance of our DSPO model, we compare it against a variety of established methods tailored for sorted portfolio tasks, with open-source implementations. These methodologies span a range of approaches, including feature engineering, time series forecasting, regression, and classification-based techniques. Specifically, we utilize LSTM models to process engineered features, adapted time series models such as DLinear\cite{zeng2023transformers} and PatchTST\cite{nie2022time} to handle extensive historical data, and employed RNN and TCN for regression tasks. Furthermore, we implemented CNNs for classification tasks, differentiating assets based on their performance tiers. Additionally, we include comparisons to other contemporary end-to-end models to highlight DSPO's unique capabilities. Detailed descriptions of these baseline methods have been relocated to the appendix to streamline the main text presentation.


\paragraph{Evaluation Metrics} We evaluate our model using both representation-based metrics (RankIC and Rank ICIR) and portfolio-based metrics (Long/LS Return and Information Ratio (IR)). RankIC measures the Spearman correlation between the predicted scores and actual returns, while RankICIR is the mean of RankICs normalized by their standard deviation over the trading period. Portfolio-based metrics, calculated using the \textit{zipline} platform\footnote{\textit{Zipline} is an open-source algorithmic trading simulator derived from Quantopian, a well-known platform in quantitative finance. It is widely used for backtesting trading strategies, enabling developers to simulate trading strategies against historical data to estimate their performance.}, including the accumulated return and Information Ratio, adjusted for commission fees and slippage rates set at 0.0002 (USD or CNY) and with volume limits and price impacts of 0.0025 and 0.01, respectively. For A-Share, only long positions were feasible due to market restrictions on short selling, while for NYSE, both long and short positions were used.

\subsection{Main Results}

\begin{table}[ht]
    \footnotesize 
    \centering
    \setlength{\tabcolsep}{2pt} 
    \caption{Comparison of performances of models under different settings.}
\begin{tabular}{@{}ccccccccc@{}}
\toprule
\multirow{2}{*}{\textbf{Formulation}}  & \multirow{2}{*}{\textbf{Model}} & \multicolumn{3}{c}{\textbf{A-Share}}        &  & \multicolumn{3}{c}{\textbf{NYSE}}                 \\ \cmidrule(lr){3-5} \cmidrule(l){7-9} 
                              &                        & RankIC(\%) & RankICIR & Return(\%) &  & RankIC(\%) & RankICIR & Longshort Return(\%) \\ \midrule
\multirow{2}{*}{Feature Eng.} & SFM                    & 1.23          & 0.29        & 19.11          &  & 1.28          & 23.2        & 25.87                \\
                              & ALSTM                  & 1.86          & 30.32        & 17.83          &  & 1.39          & 24.75        & 18.59                \\ \cmidrule(lr){2-2}
\multirow{2}{*}{Regression}   & LSTM                   & -2.66      & -0.26    & 14.99      &  & 0.94       & 0.15     & -14.18           \\
                              & TCN                    & -5.51      & -0.47    & 11.49      &  & 3.99       & 0.53     & -                \\ \cmidrule(lr){2-2}
\multirow{2}{*}{Forecasting}  & DLinear                & -0.34      & -0.08    & 9.44       &  & 0.01       & 0.04     & -75.96           \\
                              & PatchTST               & -0.35      & -0.08    & 9.44       &  & 1.26       & 0.03     & 4.27             \\ \cmidrule(lr){2-2}
Classification                & CNN                    & 4.32       & 0.48     & 14.99      &  & 3.73       & 0.76     & 59.24            \\ \cmidrule(lr){2-2}
\multirow{2}{*}{End2End}      & DIN                    & 5.51       & 0.55     & 59.35      &  & 3.41       & 0.77     & 17.44            \\
                              & DSPO (Ours)            & \textbf{9.11}       & \textbf{0.62}     & \textbf{108.74}     &  & \textbf{10.12}      & \textbf{1.05}     & \textbf{121.94}           \\ \bottomrule
\end{tabular}
    \label{tab:main}
\end{table}

The results demonstrate that our DSPO model significantly outperforms comparative models across both A-Share and NYSE markets. As shown in Tab.~\ref{tab:main}, DSPO achieved the highest RankIC and RankICIR, alongside superior portfolio returns and Information Ratio (IR).

In the A-Share market, DSPO exhibited a RankIC of 9.11\% and a RankICIR of 0.62, achieving a return of 108.74\%, substantially higher than other models. This performance underscores DSPO's effective exploitation of market characteristics, even with the constraints of long-only positions. For NYSE, DSPO maintained its superior performance, with a RankIC of 10.120\% and a RankICIR of 1.050, coupled with a long/short return of 121.94\% and an IR of 4.378. These results highlight DSPO's robustness and effectiveness in handling diverse market conditions through its end-to-end design, which integrates both high-frequency trading signals and low-frequency financial metrics more effectively than models under conventional formulation. 

\paragraph{Comparison with the Other End-to-End Method}
Th end-to-end models, DIN performed commendably in the A-Share market but demonstrated limited generalizability to the NYSE. This variance likely stems from significant differences in the input features (both high-frequency and low-frequency), time dimensions, and data distributions across markets. Such discrepancies highlight the robustness of our DSPO model in adjusting to varying market outputs and conditions, proving its superiority in dynamic financial environments.

\subsection{Case Studies}

\paragraph{Adaptability to Evolving market}
Given the fluid nature of financial markets, traditionally engineered features can become outdated, leading to degraded model performance. This was evident as feature-engineering-based approaches yielded low RankICs under our experimental setup. To test the resilience of DSPO against such signal deterioration, we incorporated the same set of long-standing feature-engineered signals as low-frequency inputs and assessed their impact. The tests are performed in the periods of 2020-2021 and 2021-2022.

\begin{table}[ht]
\centering
\caption{Ablation on high-frequency and low-frequency factors}
\begin{tabular}{@{}cccc@{}}
\toprule
                                                                    &                  & \textbf{RankIC(\%)} & \textbf{RankICIR} \\ \midrule
\multirow{3}{*}{2020-2021}                                          & LF Only          & 5.36                & 0.395             \\
                                                                    & HF Only          & 9.05                & 0.752             \\
                                                                    & HF + LF          & \textbf{11.11}      & \textbf{0.901}    \\ \cmidrule(l){2-4} 
\multirow{3}{*}{2021-2022}                                          & LF Only          & -2.16              & -0.137            \\
                                                                    & HF Only          & \textbf{9.17}      & \textbf{0.635}    \\
                                                                    & HF + LF          & 8.75               & 0.582             \\ \bottomrule
\end{tabular}
\label{tab:freq_factors}
\end{table}

As shown in Tab.~\ref{tab:freq_factors}, during the 2020 testing period, the inclusion of low-frequency, feature-engineered data proved beneficial, enhancing the DSPO model's performance. However, in 2021, this same data detracted from performance, suggesting that the predictive power of engineered features can diminish over time. Despite this, DSPO continued to achieve high RankIC scores, illustrating its ability to mitigate the adverse effects of deteriorating signals significantly, thus maintaining its efficacy in volatile market conditions.

\paragraph{Stability Across Different Trials}
Running multiple experiments with different seeds and employing model ensembles are typical approaches in conventional settings. This practice is often necessitated by the inherent instability and significant performance variance observed across different runs. To investigate whether DSPO is similarly affected, we conducted eight trials using the same hyperparameter settings but varying only the seed. These trials were compared with other end-to-end and classification settings.

As shown in Fig.~\ref{fig:mesh1}, DSPO consistently outperforms the other models in terms of cumulative return during the backtesting period, as illustrated by the red line. DSPO not only achieves higher returns but also demonstrates exceptional stability across different runs, evidenced by the tightly clustered variance (shaded red area). This contrasts markedly with DIN (shown in green), which, while securing second place in cumulative returns, exhibits considerably greater variance, indicating less reliability. The CNN models (represented in blue), despite achieving lower returns, show intermediate variance, suggesting moderate stability but underperformance relative to DSPO. These results underscore DSPO’s capability to deliver robust performance, potentially reducing the need for model ensembles during inference phases, thereby simplifying model deployment and maintenance.

\subsection{Ablation Studies}

\paragraph{Effectiveness of DSPO Modules}
To ascertain the contribution of individual components within the DSPO model, we conducted detailed ablation studies. Our findings confirm that each module significantly enhances the overall performance. Notably, the Monotonic Logarithmic Regression (MonLR) loss facilitated more effective optimization, while the Multi-Frequency Fusion module proved crucial for modeling the interdependencies among individual stocks, as illustrated in Tab.~\ref{tab:modules_abl}.

\begin{table}[ht]
    \centering
    \centering
    \caption{Ablation on DSPO Modules}
    \begin{tabular}{@{}lcc@{}}
\toprule
\multicolumn{1}{c}{\textbf{Model}} & \textbf{RankIC(\%)} & \textbf{RankICIR} \\ \midrule
DSPO                               & \textbf{9.11}   & \textbf{0.62}     \\
w/o MonLR Loss                     & 6.57            & 0.76              \\
w/o Multi-Stock                    & 3.73            & 0.76              \\ \bottomrule
\end{tabular}
\label{tab:modules_abl}
\end{table}

\paragraph{Impact of Sub-Sampling Strategy}
An ablation study on the sub-sampling strategy was performed to evaluate the effect of the number of sub-sampled stocks. As depicted in Tab.~\ref{tab:subsample}, we found that, with sub-subsampled cross-sections with sufficient number of stocks, RankIC and ICIR are maintained in a high-performance level, suggesting that the convergence to the expectation of the objective of directly optimizing the ranking order future returns. If the number of stocks is too little, the sub-sampled cross-sections cannot represent the characteristic of the original cross-section, leading to satisfactory results.
\begin{table}[ht]
    \centering
    \centering
    \caption{Ablation on Num. of Sub-Sampled Stocks}
\begin{tabular}{@{}ccc@{}}
\toprule
\textbf{Num. Stocks} & \textbf{RankIC(\%)} & \textbf{RankICIR} \\ \midrule
100                  & 6.67  & 0.62     \\
500                  & 9.92            & 0.90              \\
1500                 & 11.11           & 0.90              \\
2000                 & 9.89            & 0.76              \\ \bottomrule
\end{tabular}
    \label{tab:subsample}
\end{table}


\paragraph{On Initial Capital on Performance}
The scale of investment significantly influences performance outcomes. To understand this effect, we conducted ablation studies on DSPO with varying levels of initial capital and evaluated their impact on various returns and risk metrics within the NYSE context. As illustrated in ~\ref{tab:capital_ablation}, we observe a decline in the accumulative returns as the initial capital increases, in both long-only and long-short settings, Correspondingly, the Information Ratio (IR) and Maximum Drawdown (MDD) ratio also deteriorate.

\begin{table}[H]
    \centering
    \caption{Ablation study on initial capital}
\begin{tabular}{@{}cccclccc@{}}
\toprule
\multirow{2}{*}{\textbf{Investment(\$)}} & \multicolumn{3}{c}{\textbf{Long}}                    & \textbf{} & \multicolumn{3}{c}{\textbf{Long-Short}}              \\ \cmidrule(lr){2-4} \cmidrule(l){6-8} 
                                         & \textbf{Return(\%)} & \textbf{IR} & \textbf{MDD(\%)} & \textbf{} & \textbf{Return(\%)} & \textbf{IR} & \textbf{MDD(\%)} \\ \midrule
10K                                      & 94.05               & 2.98        & -6.18            &           & 157.83              & 2.90        & -1.55            \\
100K                                     & 115.06              & 3.41        & -6.09            &           & 121.94              & 2.90        & -1.47            \\
1M                                       & 50.14               & 3.09        & -2.93            &           & 55.52               & 2.66        & -0.82            \\
10M                                      & 9.18                & -3.79       & -0.45            &           & 12.57               & -3.51       & -0.41            \\ \bottomrule
\end{tabular}
    \label{tab:capital_ablation}
\end{table}

This trend suggests that while DSPO is effective in accurately predicting rank orders, it does not account for other critical factors such as market liquidity and risk exposure. These findings underscore the need for strategies that are tailored to different investment scales, emphasizing the integration of risk management components to optimize performance across varying capital allocations.

\section{Limitations and Future Work}
\label{sec:limiation}

Our findings highlight some limitations in DSPO’s current implementation, particularly its insensitivity to changes in investment scale and associated market dynamics, such as liquidity and risk exposure. The observed decline in performance metrics, like the Information Ratio and Maximum Drawdown, as investment capital increases underscores a critical shortfall in the model’s adaptability to varying market conditions. This limitation is especially concerning in scenarios involving substantial capital where the impact of market liquidity and volatility is more pronounced.

To address these limitations, future research should focus on integrating risk management strategies into the DSPO framework. Developing adaptive algorithms that can dynamically adjust to the size of the capital and prevailing market conditions could enhance the robustness and applicability of DSPO across different investment scales. Additionally, incorporating factors such as market liquidity, volatility, and other macroeconomic indicators could provide a more holistic approach to predictive modeling in financial markets.

\section{Conclusion}

This study presents Direct Sorted Portfolio Optimization, a novel end-to-end framework for constructing characteristic-sorted portfolios by integrating high-frequency market data with daily-frequency fundamental data. Through a strategic sub-sampling approach and the introduction of Monotonical Logistic Regression, our model demonstrates improved robustness and efficiency. DSPO significantly outperforms state-of-the-art methods in RankIC, portfolio returns, and risk-adjusted performance metrics. Our results underscore DSPO’s superior capacity to leverage diverse market data and its robustness in managing the complexities of modern financial markets, paving the way for future advancements in quantitative investment strategies.

\newpage
\bibliographystyle{ACM-Reference-Format}
\bibliography{ref}

\newpage
\appendix

\section{Impact of Sub-Sampling Stradegy}
The crux of our sub-sampling strategy lies in the intra-day cross-sector selection, which fundamentally serves as a technique for lossless data augmentation. By selectively sampling both temporal (day-wise) and spatial (ticker-wise) dimensions of our dataset, we effectively augment our training data without the introduction of noise or synthetic data.

\paragraph{Intra-Day Sub-Sampling}
This technique involves randomly selecting \( m \) distinct trading days from our dataset. The selection is not sequential but rather stochastic, ensuring coverage across various market conditions without a temporal bias. Each selected day is treated as a distinct sample, potentially representing a unique market scenario.

\paragraph{Cross-Sector Sub-Sampling}
In addition to day-wise sampling, we further refine our data augmentation by selecting \( k \) stocks randomly from each of the \( m \) trading days' active stock pool. This approach ensures sectorial diversity within our samples and allows the model to learn generalized features across various market segments. The process of cross-sector sub-sampling is independent for each trading day, further increasing the diversity of the training set.

\textbf{Combating Data Scarcity and Overfitting}\
The aforementioned sub-sampling strategy substantially increases the number of unique training samples we can extract from a limited dataset. For a given trading day with \( n \) active stocks, the number of possible training samples (\( C_{n}^{k} \)) that can be generated by selecting \( k \) stocks can be mathematically represented as:

\begin{equation}
    C_{n}^{k} = \binom{n}{k} = \frac{n!}{k!(n-k)!}
\end{equation}

For example, considering \( n = 4000 \) stocks in a trading day and sampling \( k = 1000 \) stocks, the number of potential unique combinations is:

\begin{equation}
    \binom{4000}{1000} \approx 1.095 \times 10^{975}
\end{equation}

This combinatorial explosion of training samples mitigates the risk of overfitting by exposing the model to a wider spectrum of the data distribution, fostering robustness and enhancing generalizability to unseen market data.

\section{Dataset Detail}
The comprehensive datasets comprising both high-frequency and low-frequency financial data offer a rich array of metrics, meticulously derived from market transactions and financial reports of all actively tradable stocks inside the same markets. These data sets essentially represent raw stock data. High-frequency data, gathered from minute-to-minute or even second-to-second transactions on the NYSE, adeptly captures the rapid and dynamic fluctuations within the stock market. In contrast, low-frequency data is compiled from more aggregated sources, such as daily, weekly, or monthly summaries of earnings reports, industry analyses, and economic indicators. Collectively, these datasets provide an essential foundation for a thorough understanding of market behaviors and financial standings, crucial for robust quantitative finance analysis. The majority of feature engineering approaches construct the input feature for predictive model from these raw stock data. We are actively processing this dataset for future release.

\subsection{NYSE}

\paragraph{High-frequency data}
The dataset comprises a comprehensive collection of high-frequency trading data from the New York Stock Exchange (NYSE) covering the years 2019 to 2024. It is segmented into several key metrics that are essential for financial analysis and trading strategy development:

\begin{itemize}
    \item \textbf{Closing Prices} - This data captures the final trading prices of stocks at the close of each trading day, providing insights into the end-of-day valuation and the settling point of market activities.

    \item \textbf{High Prices} - Records the highest trading prices of stocks during the trading sessions, reflecting the peak valuation reached within the day and serving as a marker of maximum intra-day volatility.

    \item \textbf{Low Prices} - Comprises the lowest prices at which stocks were traded during the day, offering a perspective on the trough or minimum market valuation and intra-day price drops.

    \item \textbf{Opening Prices} - Includes the initial prices at which stocks start trading at the beginning of the market day, crucial for assessing the starting valuation and early market trends.

    \item \textbf{Trading Volume} - This metric indicates the total number of stocks traded during the day, which is vital for understanding market liquidity and the intensity of trading activity.

    \item \textbf{Volume-Weighted Average Prices} - Provides the average price of stocks traded throughout the day, weighted by their trading volume. This measure is particularly useful for assessing the overall market performance and the execution efficiency of trades.
\end{itemize}

\paragraph{Low-Frequency Data}


This dataset encompasses a broad range of low-frequency financial data essential for comprehensive market analysis and evaluation of investment opportunities. Each category of data offers unique insights into market trends, company performance, and economic factors over extended periods:

\begin{itemize}
    \item \textbf{Average Price} - Tracks the mean trading price of stocks, offering insights into overall market valuation trends over daily periods.

    \item \textbf{Daily Turnover Rate} - Represents the ratio of shares traded compared to the total shares available, indicating the liquidity and investor activity on a daily basis.

    \item \textbf{Daily Volume} - Records the total number of shares traded each day, serving as a direct indicator of market activity and interest.

    \item \textbf{Daily Yield} - Details the daily return on investments for stocks, crucial for analyzing day-to-day performance and profitability.

    \item \textbf{Industry Classification} - Categorizes companies into various industries, facilitating sector-specific analysis and benchmarking.

    \item \textbf{Market Capitalization} - Lists the total market value of publicly traded shares, providing a measure of company size and market dominance.

    \item \textbf{Monthly Turnover Rate} - Measures monthly liquidity and trading activity, offering a longer-term perspective compared to daily rates.

    \item \textbf{Price-to-Book Ratio} - Compares a company’s market price to its book value, helping assess whether a stock is undervalued or overvalued.

    \item \textbf{Price-to-Earnings Ratio} - Indicates how much investors are willing to pay per dollar of earnings, a key metric for valuation.

    \item \textbf{Price/Earnings to Growth Ratio} - Refines the P/E ratio by considering earnings growth, enhancing valuation accuracy.

    \item \textbf{Weekly Turnover Rate} - Provides insights into weekly trading dynamics, balancing between daily fluctuations and monthly trends.

    \item \textbf{Weekly Yield} - Offers a summary of weekly investment returns, useful for short-term performance assessment.
\end{itemize}

\subsection{A-shares Market}

\paragraph{High-frequency data}

\begin{itemize}
    \item \textbf{Closing Prices} - Captures the final trading prices of stocks at the close of each trading day, offering insights into the end-of-day valuation and market settling points.

    \item \textbf{High Prices} - Records the highest trading prices of stocks during the sessions, reflecting peak valuations and maximum intra-day volatility.

    \item \textbf{Low Prices} - Comprises the lowest prices at which stocks were traded during the day, providing insights into the minimum market valuations and intra-day price fluctuations.

    \item \textbf{Opening Prices} - Includes the initial prices at which stocks start trading at the beginning of the market day, crucial for assessing early market trends and starting valuations.

    \item \textbf{Trading Volume} - Indicates the total number of stocks traded during the day, essential for understanding market liquidity and the intensity of trading activity.

    \item \textbf{Volume-Weighted Average Prices} - Provides the average price at which stocks traded throughout the day, weighted by their trading volume. This measure is particularly valuable for assessing overall market performance and trade execution efficiency.
\end{itemize}

These data points collectively facilitate a comprehensive analysis of the A-shares market, enabling investors and analysts to track market trends, evaluate stock performances, and develop effective trading strategies.

\paragraph{Low-Frequency Data}
The dataset also includes a diverse range of low-frequency data which provides deeper insights into the financial standing and market behavior of A-shares over longer periods:

\begin{itemize}
    \item \textbf{Average Price} - Provides a mean trading price of stocks, crucial for understanding broader market valuation trends over time.

    \item \textbf{Closing Prices} - Offers data on the final trading prices at the close of each trading day, essential for end-of-day market analysis.

    \item \textbf{Daily Yield} - Details the daily return on investments, highlighting day-to-day performance and profitability of stocks.

    \item \textbf{Dividend Yields} - Shows the percentage of earnings distributed to shareholders, important for assessing the income-generating potential of investments.

    \item \textbf{Free Float Shares} - Indicates the number of shares available for trading, key for understanding market liquidity.

    \item \textbf{High/Low Prices} - Records the highest and lowest prices stocks reached during trading sessions, reflecting intra-day volatility.

    \item \textbf{Market Value} - Lists the total market capitalization of companies, a fundamental indicator of company size and market share.

    \item \textbf{Monthly Turnover Rate} - Measures the frequency of stock trading, useful for gauging market activity and investor interest over each month.

    \item \textbf{Price Metrics (PB, PE, PEG, PS)} - Various valuation metrics, such as Price-to-Book, Price-to-Earnings, Price/Earnings to Growth, and Price-to-Sales ratios, provide comprehensive insights into the financial worth and growth potential of stocks.

\end{itemize}

\section{Detailed Experimental Settings}
In the section, we explain our experimental setup in detail.

\subsection{Detailed Settup for Other computation methods}
We employ several distinct approaches to model and analyze financial data, each leveraging unique models and computational strategies:

\begin{itemize}
    \item \textbf{Feature Engineering Based:} High-frequency market data is transformed into multiple predictive signals through a Long Short-Term Memory (LSTM) model. This model uses a Mean Squared Error (MSE) loss function to regress returns, focusing on capturing the temporal dependencies within the data.
    
    \item \textbf{Time Series Forecasting Based:} For time series forecasting, we utilize DLinear and PatchTST models, tailored to our dataset's requirements. We perform downsampling to manage the computational load, setting the lookback window to 720 and the forecasting horizon to 144, enabling more effective long-term prediction.
    
    \item \textbf{Regression Based:} Recurrent Neural Network (RNN) and Temporal Convolutional Network (TCN) are deployed to process high-frequency signals. These models also employ an MSE loss, optimized for continuous output prediction, which is crucial for tracking subtle fluctuations in high-frequency data.
    
    \item \textbf{Classification Based:} Assets are classified into three categories: top, middle, and bottom deciles within their respective cross-sectors. A Convolutional Neural Network (CNN) trained with cross-entropy loss calculates the probability distributions over these categories. The final scores are obtained by contrasting the logistic scores of the top and bottom deciles, thus highlighting the most and least promising assets.
    
    \item \textbf{Other Previous End-to-End Formulations:} Following the framework outlined in \cite{liu2023deep}, we adapt these methodologies specifically for intraday trading. The evaluation relies on using returns from the past weeks to compute a loss based on the Sharpe ratio, as mentioned in the Direct Inference Network (DIN), enhancing our model's focus on risk-adjusted returns.
\end{itemize}

\subsection{Detailed Metrics Computation}

To provide a thorough understanding of the metrics used in evaluating the Direct Sorted Portfolio Optimization (DSPO) model, we detail the computational methods and formulas for each key metric below. These metrics include RankIC, accumulated returns, and variance, which are essential for assessing the performance and stability of the model across different market conditions.

\paragraph{Rank Information Coefficient (RankIC)}
The Rank Information Coefficient (RankIC) is a measure of the rank correlation between predicted and actual future returns. It is calculated using the Spearman's rank correlation coefficient, which assesses how well the relationship between two variables can be described using a monotonic function. The formula for RankIC is given by:

\begin{equation}
    \text{RankIC} = \rho_{R(X), R(Y)} = \frac{\text{cov}(R(X), R(Y))}{\sigma_{R(X)} \sigma_{R(Y)}}
\end{equation}

Where \(R(X)\) is the predicted return and \(R(Y)\) is the ground truth return of the entire corss-sector.

\paragraph{Information Coefficient Information Ratio (ICIR)}
The Information Coefficient Information Ratio (ICIR) is an extension of the Rank Information Coefficient (RankIC) and measures the consistency of predictive skill. It is calculated as the mean of the RankIC values over the trading period, normalized by their standard deviation:

\begin{equation}
    \text{ICIR} = \frac{\text{Mean of RankICs}}{\text{Standard Deviation of RankICs}}
\end{equation}

This metric provides an understanding of how stable the RankIC values are over time, offering insights into the reliability of the model’s predictions.

\paragraph{Maximum Drawdown (MDD)}
Maximum Drawdown (MDD) is a measure of the largest single drop from peak to trough in the value of a portfolio, before a new peak is achieved. It is an indicator of downside risk over a specified time period. MDD is defined as:

\begin{equation}
    \text{MDD} = \max_{\tau \in (t, T)} \left( \max_{t \in (0, \tau)} P_t - P_\tau \right)
\end{equation}

where \(P_t\) is the portfolio value at time \(t\), and \(T\) is the total period considered. MDD assesses the largest loss that would have been experienced by an investor in the period.

\paragraph{Information Ratio (IR)}
The Information Ratio (IR) measures the return of a portfolio relative to the return of a benchmark index, adjusted for the volatility of the portfolio returns. It is computed as the ratio of the excess return of the portfolio to the standard deviation of these excess returns:

\begin{equation}
    \text{IR} = \frac{\text{Mean of Excess Returns}}{\text{Standard Deviation of Excess Returns}}
\end{equation}

where excess returns are the portfolio returns minus the benchmark returns. IR is particularly useful for evaluating the risk-adjusted performance of a portfolio against its benchmark.

\paragraph{Calculation in Zipline}
Metrics such as the accumulated return and Information Ratio are calculated using the \textit{zipline} platform. Adjustments for commission fees and slippage rates are incorporated into these calculations, with fees set at 0.0002 (Dollar or Yuan) and slippage modeled with volume limits and price impacts of 0.0025 and 0.01, respectively. These adjustments ensure that the metrics reflect realistic trading scenarios, accounting for the cost and market impact of transactions.

These detailed metrics, computed as described, provide a comprehensive evaluation of the model's performance, encompassing aspects of accuracy, consistency, risk, and return relative to a benchmark. By applying these metrics, we can thoroughly assess the robustness and effectiveness of the investment strategies implemented through our model.

\subsection{Hyperparameter Selection for Training Experiments}

Our experimental setup employs default hyperparameters to establish a consistent baseline for evaluating the performance of the MSTransformer model in financial time series forecasting. Key hyperparameters include:

\begin{itemize}
    \item \textbf{Sample Size:} The number of samples per batch is set to 1000, allowing for sufficient data representation while managing computational load.
    \item \textbf{Batch Size:} A batch size of 6 is used to ensure that each batch is representative of the overall dataset without overwhelming the GPU memory.
    \item \textbf{Number of Epochs:} The model is trained for 80 epochs, balancing between adequate learning time and preventing overfitting.
    \item \textbf{Learning Rate:} Set to $1 \times 10^{-5}$, this low learning rate helps in fine-tuning the model's weights with precision, especially important in the noisy financial data context.
    \item \textbf{Gradient Clipping:} The gradient clipping value is set to 0.5 with a norm clipping algorithm to prevent exploding gradients, a common issue with LSTM networks.
    \item  \textbf{Optimizer:}
\end{itemize}

\paragraph{Optimizer and Learning Rate Scheduling}
For the optimization of our model, we use the AdamW optimizer, known for its effectiveness in handling sparse gradients and improving generalization by decoupling the weight decay from the gradient updates. The specific parameters for AdamW include:
\begin{itemize}
    \item \textbf{Learning Rate (lr):} Initially set to $1 \times 10^{-5}$.
    \item \textbf{Betas:} Set to (0.9, 0.98) to control the exponential decay rates for the moment estimates.
    \item \textbf{Epsilon (eps):} Set to a very low value of $1 \times 10^{-9}$ to improve numerical stability.
\end{itemize}

Additionally, we implement a learning rate scheduler based on the Noam scheme, which is particularly useful for training deep learning models with transformer architectures. The Noam scheduler adjusts the learning rate according to the formula:
\begin{equation}
    lr = \text{scale} \cdot \text{model\_size}^{-0.5} \cdot \min(\text{step}^{-0.5}, \text{step} \cdot \text{warmup\_steps}^{-1.5})
\end{equation}
where \textit{scale} is a constant factor, \textit{model\_size} is the total number of trainable parameters, \textit{step} is the current step, and \textit{warmup\_steps} is the number of steps during the warm-up period. This method allows for an initial increase in learning rate to accelerate the model's early learning phase, followed by a gradual decay to fine-tune model parameters.

Model performance is monitored using the Ranked Information Coefficient (IC), and we select the checkpoint with the second best for test. All of our experiments are performed using A800 80G gpus.

\subsection{Detailed Featured Engineering Procedure for Corresponding Models}

We follow the procedure explained in \cite{haitong_performance_nodate} to map high-frequency stock data to low-frequency ones for the corresponding baselines.

\paragraph{Realized Variance (RVar)}
Realized Variance is computed using intra-day returns, capturing the total variance within the trading day:
\begin{equation}
    RVar_i = \sum_{j=1}^N r_{ij}^2
\end{equation}
where \(r_{ij}\) represents the return for interval \(j\) on day \(i\), and \(N\) is the number of intervals in a day. This metric is pivotal for assessing the volatility of the asset within a single trading day.

\paragraph{Realized Skewness (RSkew)}
Realized Skewness measures the asymmetry of the return distribution around its mean:
\begin{equation}
    RSkew_i = \frac{\sqrt{N}\sum_{j=1}^N r_{ij}^3}{(RVar_i)^{3/2}}
\end{equation}
It helps identify the direction and strength of the tails of the return distribution relative to a normal distribution.

\paragraph{Realized Kurtosis (RKurt)}
Realized Kurtosis quantifies the tailedness of the return distribution:
\begin{equation}
    RKurtosis_i = \frac{N\sum_{j=1}^N r_{ij}^4}{(RVar_i)^2}
\end{equation}
This statistic indicates whether the returns have heavier or lighter tails than a normal distribution, which can be critical for risk management.

\paragraph{Downside Risk Beta}
Downside Risk Beta is a statistical measure designed to capture the sensitivity of an asset's returns to movements in a benchmark during periods when the benchmark's returns are declining. This measure provides a nuanced view of an asset's risk profile, especially during market downturns. The Downside Risk Beta (\( \beta^- \)) is calculated using the formula:
\begin{equation}
    \beta^- = \frac{\sum_{j=1}^N r_{ij}^2 \cdot \mathbf{1}_{r_{ij} < 0}}{\sum_{j=1}^N r_{ij}^2}
\end{equation}
where \( r_{ij} \) represents the return for interval \( j \) on day \( i \), \( N \) is the number of intervals in a day, and \( \mathbf{1}_{r_{ij} < 0} \) is an indicator function that takes the value 1 if \( r_{ij} \) is less than zero (indicating a negative return) and 0 otherwise.

\paragraph{Flow-In Ratio}
The Flow-In Ratio is designed to measure the relative buying or selling pressure during the trading day, offering insights into market sentiment and liquidity dynamics. The Flow-In Ratio is calculated using the formula:
\begin{equation}
    \text{FlowInRatio} = \frac{\sum_{i,j} \left( \text{Volume}_{ij} \cdot \text{Close}_{ij} \cdot \frac{\left| \text{Close}_{ij} - \text{Close}_{ij-1} \right|}{\sum_i \text{Amount}_{i,\text{total}}} \right)}{\sum_{i,j} \left| \text{OI}_{ij} - \text{OI}_{ij-1} \right| \cdot \text{Close}_{ij}}
\end{equation}
where \( \text{Volume}_{ij} \) is the traded volume at interval \( j \) on day \( i \), \( \text{Close}_{ij} \) is the closing price at interval \( j \), and \( \text{Amount}_{i,\text{total}} \) is the total trading amount on day \( i \).

\paragraph{Trend Strength}
Trend Strength is a metric used to assess the robustness of a price trend over a specified period. It is calculated as follows:
\begin{equation}
    \text{TrendStrength} = \frac{P_n - P_1}{\sum_{i=2}^{n} \left| P_i - P_{i-1} \right|}
\end{equation}
where \( P_t \) represents the price at time \( t \), \( P_1 \) is the price at the start of the period, and \( P_n \) is the price at the end of the period.

These statistical measures are then aggregated or sampled appropriately to match the input frequency required by our models, which is daily frequency. Although these feature engineer approaches were shown to be effective a few years ago by \cite{haitong_performance_nodate}, they are long established and outdated during our backtesting periods.

\newpage
\newpage
\section*{NeurIPS Paper Checklist}

The checklist is designed to encourage best practices for responsible machine learning research, addressing issues of reproducibility, transparency, research ethics, and societal impact. Do not remove the checklist: {\bf The papers not including the checklist will be desk rejected.} The checklist should follow the references and precede the (optional) supplemental material. The checklist does NOT count towards the page
limit.

\begin{enumerate}

\item {\bf Claims}
    \item[] Question: Do the main claims made in the abstract and introduction accurately reflect the paper's contributions and scope?
    \item[] Answer: \answerYes{} 
    \item[] Justification: First, we first illustrate our motivation, then roll out the detial description in the methodology section. Lastly, we conduct various experiments to support our claims.

\item {\bf Limitations}
    \item[] Question: Does the paper discuss the limitations of the work performed by the authors?
    \item[] Answer: \answerYes{} 
    \item[] Justification: We first introduce our limitation with experimental results in section~\ref{experiments}, then state the limitations in section~\ref{sec:limiation}

\item {\bf Theory Assumptions and Proofs}
    \item[] Question: For each theoretical result, does the paper provide the full set of assumptions and a complete (and correct) proof?
    \item[] Answer: \answerYes{} 
    \item[] Justification: We propse our assumption in section ~\ref{related_work} and section ~\ref{sec:monLR} provide the derivation.

    \item {\bf Experimental Result Reproducibility}
    \item[] Question: Does the paper fully disclose all the information needed to reproduce the main experimental results of the paper to the extent that it affects the main claims and/or conclusions of the paper (regardless of whether the code and data are provided or not)?
    \item[] Answer: \answerYes{} 
    \item[] Justification: We disclose all the information in experimental settup in experiment section and appendix. We will release our code for reproducibility.

\item {\bf Open access to data and code}
    \item[] Question: Does the paper provide open access to the data and code, with sufficient instructions to faithfully reproduce the main experimental results, as described in supplemental material?
    \item[] Answer: \answerYes{} 
    \item[] Justification: We will release the our data and code after acceptances.

\item {\bf Experimental Setting/Details}
    \item[] Question: Does the paper specify all the training and test details (e.g., data splits, hyperparameters, how they were chosen, type of optimizer, etc.) necessary to understand the results?
    \item[] Answer: \answerYes{} 
    \item[] Justification: See Section ~\ref{experiments} and Appendix.

\item {\bf Experiment Statistical Significance}
    \item[] Question: Does the paper report error bars suitably and correctly defined or other appropriate information about the statistical significance of the experiments?
    \item[] Answer: \answerYes{} 
    \item[] Justification: ~\ref{fig:mesh1} document, report and compare the variance across differnece trials for different approaches.

\item {\bf Experiments Compute Resources}
    \item[] Question: For each experiment, does the paper provide sufficient information on the computer resources (type of compute workers, memory, time of execution) needed to reproduce the experiments?
    \item[] Answer: \answerYes{} 
    \item[] Justification: We report type of compute workers in Appendix.
    \item[] Guidelines:
    
\item {\bf Code Of Ethics}
    \item[] Question: Does the research conducted in the paper conform, in every respect, with the NeurIPS Code of Ethics \url{https://neurips.cc/public/EthicsGuidelines}?
    \item[] Answer: \answerYes{} 
    \item[] Justification: To our knowledge, our paper follows the NeurIPS Code of Ethics

\item {\bf Broader Impacts}
    \item[] Question: Does the paper discuss both potential positive societal impacts and negative societal impacts of the work performed?
    \item[] Answer: \answerYes{} 
    \item[] Justification: We discussed the social impact of optimizing for sorted portfolio.

\item {\bf Safeguards}
    \item[] Question: Does the paper describe safeguards that have been put in place for responsible release of data or models that have a high risk for misuse (e.g., pretrained language models, image generators, or scraped datasets)?
    \item[] Answer: \answerNA{} 
    \item[] Justification: 

\item {\bf Licenses for existing assets}
    \item[] Question: Are the creators or original owners of assets (e.g., code, data, models), used in the paper, properly credited and are the license and terms of use explicitly mentioned and properly respected?
    \item[] Answer: \answerYes{} 
    \item[] Justification: we have properly cited all the related papers that are used or compared within our paper.

\item {\bf New Assets}
    \item[] Question: Are new assets introduced in the paper well documented and is the documentation provided alongside the assets?
    \item[] Answer: \answerYes{} 
    \item[] Justification: see our Appendix, we will release our code and model after acceptance.

\item {\bf Crowdsourcing and Research with Human Subjects}
    \item[] Question: For crowdsourcing experiments and research with human subjects, does the paper include the full text of instructions given to participants and screenshots, if applicable, as well as details about compensation (if any)? 
    \item[] Answer: \answerNA{} 
    \item[] Justification: our paper does not involve crowdsourcing nor research with human subjects.

\item {\bf Institutional Review Board (IRB) Approvals or Equivalent for Research with Human Subjects}
    \item[] Question: Does the paper describe potential risks incurred by study participants, whether such risks were disclosed to the subjects, and whether Institutional Review Board (IRB) approvals (or an equivalent approval/review based on the requirements of your country or institution) were obtained?
    \item[] Answer: \answerNA{} 
    \item[] Justification: our paper does not involve crowdsourcing nor research with human subjects.

\end{enumerate}

\end{document}